# The effect of the dispersal kernel on isolation-by-distance in a continuous population.

Tara N. Furstenau and Reed A. Cartwright[*]

The Biodesign Institute and School of Life Sciences

Arizona State University, Tempe, AZ 85287-5301 USA

[*]cartwright@asu.edu

January 5, 2015

**Abstract**

Under models of isolation-by-distance, population structure is determined by the probability of identity-by-descent between pairs of genes according to the geographic distance between them. Well established analytical results indicate that the relationship between geographical and genetic distance depends mostly on the neighborhood size of the population, $N_b = 4\pi\sigma^2 D_e$, which represents a standardized measure of dispersal. To test this prediction, we model local dispersal of haploid individuals on a two-dimensional torus using four dispersal kernels: Rayleigh, exponential, half-normal and triangular. When neighborhood size is held constant, the distributions produce similar patterns of isolation-by-distance, confirming predictions. Considering this, we propose that the triangular distribution is the appropriate null distribution for isolation-by-distance studies. Under the triangular distribution, dispersal is uniform within an area of $4\pi\sigma^2$ (i.e. the neighborhood area), which suggests that the common description of neighborhood size as a measure of a local panmictic population is valid for popular families of dispersal distributions. We further show how to draw from the triangular distribution efficiently and argue that it should be utilized in other studies in which computational efficiency is important.

# 1 Introduction

For many populations, individuals do not exist in discrete patches or demes; instead they are spread across a continuous landscape. Although there are no barriers separating individuals, dispersal distances are often limited, and individuals that are near one another in space will be more similar genetically than individuals further apart. This phenomenon is known as isolation-by-distance and introduces a spatial component that should be considered when studying population genetic processes (Jongejans *et al*, 2008). Unfortunately, incorporating multiple dimensions of space at fine scales into analytical models is often analytically intractable (Epperson *et al*, 2010). Therefore, many researchers have turned to spatially-explicit, individual-based computer simulations which offer a more flexible way to incorporate spatial complexity into biological models (e.g. Barton *et al*, 2013; Cartwright, 2009; Epperson, 2003; Novembre and Stephens, 2008; Rousset, 2004; Slatkin, 1993).

The strength of isolation-by-distance can be quantified by neighborhood size, $N_b = 4\pi\sigma^2 D$, where $D$ is the effective density of individuals, and $2\sigma^2$ is the mean squared distance of dispersal, i.e. non-central second moment of Euclidean parent-offspring distance (Barton *et al*, 2013; Epperson, 2007; Malécot, 1969; Rousset, 1997, 2004; Wright, 1946). Neighborhood size is important because it determines the inverse relationship between distance and genetic similarity (Rousset, 2000; Barton *et al*, 2013) and conceptually measures the effective number of individuals found in a circle of radius $2\sigma$.

Ideally, when modeling isolation-by-distance, the dispersal distribution would be selected based on how well it fits the dispersal kernel estimated from natural populations. However, theory predicts that the actual dispersal distribution does not have a significant effect on the evolution of isolation-by-distance. Therefore, it should be possible to select any dispersal distribution as long as $\sigma^2$ stays constant. Unfortunately, Rousset (1997) points out that the relationship is more complicated and does depends on the shape of the dispersal distribution, but, for many classes of distributions, $\sigma^2$ is likely the only parameter that matters.

Here we develop a spatially-explicit, individual-based model to simulate local dispersal in a continuous population to determine if patterns of isolation-by-distance vary based on the shape of four different dispersal distributions: Rayleigh, half-normal, exponential, and triangular. Each dispersal distribution has a different shape but they can be parameterized such that their non-central second moment is $2\sigma^2$. If the simulations reveal a similar pattern of isolation-by-distance across all four dispersal distributions, we can conclude, at least for these distributions, that $\sigma^2$ is the main determining factor of genetic differentiation with distance. Consequently, when designing isolation-by-distance simulations, researchers can safely choose a dispersal distribution based on computational needs instead of biological fit.

## 2 Methods

### 2.1 Simulation

In our model, a population exists on a $N_i \times N_j$ rectangular lattice with periodic boundaries (a torus), and individuals are uniformly spaced with a single individual per cell. Individuals contain a single, haploid, neutral genetic locus. The simulation is initialized with $N_i \times N_j$ unique haplotypes. Generations are discrete, and individuals reproduce by generating clonal offspring which disperse from the maternal cell according to a chosen distribution. Offspring experience mutations according to the infinite alleles model at rate $\mu$.

As offspring land on their destination cell they are immediately accepted or rejected using a reservoir sampling method which allows us to uniformly sample offspring at each location as they arrive instead of storing them all in memory (Vitter, 1985). When dispersal is complete, there will be a maximum of one offspring per cell and that offspring will become a parent in the next generation. While testing the simulation, we determined that if each parent generated 15 offspring, the number of empty cells per generation was negligible.

### 2.2 Modeling Dispersal

In our spatially-explicit model, space is represented on a discrete lattice, and therefore, we approximate continuous dispersal distributions by discretization distances on the lattice (Chesson and Lee, 2005; Chipperfield *et al*, 2011). We use four dispersal kernels (Table 1): $f(d, \alpha; \sigma)$, where $\alpha \in [0, 2\pi]$ is a uniformly distributed direction, and $d > 0$ is the distance of the dispersal (i.e. polar coordinates). As described above, $\sigma$ is the square root of one-half the second-moment of dispersal distance. Once the angle and distance are drawn, the final position is determined by converting the polar coordinates into rectangular coordinates and adding them to the parent's position. The new coordinates are then rounded to determine the cell in which they fall. This is similar to the centroid-to-area approximation of continuous dispersal kernels method described by Chipperfield *et al* (2011) which showed minimal deviation from expectations under continuous dispersal.

The Rayleigh and exponential distributions are two popular distributions in isolation-by-distance simulations. The Rayleigh is used because it can be modeled as independent normal distributions on the horizontal and vertical axes. The exponential distribution is a leptokurtic dispersal kernel with a sharp peak at the origin and a fatter tail. This distribution is relevant because natural dispersal is typically leptokurtic (Clark, 1998). The half-normal distribution is equivalent to a normal distribution that has been folded over the y-axis and returns the absolute value of normally distributed distances. The triangular distribution is defined on the range $d \in [0, 2\sigma]$ and returns distances such that positions are uniformly sampled from a circle with area $4\pi\sigma^2$. While the triangular distribution is an unusual choice for isolation-by-distance simulations, it has some interesting properties that will be discussed below.

Because the dispersal function is executed over 100-billion times per simulation, it is important that the implementation be as efficient as possible. We use an efficient xorshift algorithm for uniform pseudo-random number generation and the ziggurat rejection sampling algorithm to draw values from exponential, and half-normal distributions (Marsaglia and Tsang, 2000; Marsaglia, 2003). The ziggurat algorithm is a very fast algorithm which relies on precomputed look-up tables and a minimal number of computations. This algorithm can be applied to monotone decreasing probability distributions and symmetric unimodal distributions. Dispersal using the Rayleigh distribution can be generated quickly by simulating pairs of horizontal and vertical offsets from independent normal distributions. Rayleigh dispersal is further optimized by using the ziggurat algorithm for the normal distribution.

The bounded range of the triangular distribution allows us to simulate dispersal efficiently. We bypass the costly conversions from polar to Cartesian coordinates by using a discrete sampling method. This method requires some complicated pre-processing but it allows us to draw relative dispersal positions in constant time. In two-dimensions, the triangular distribution produces a uniform distribution over a disk. Given the $\sigma$ parameter, we can calculate the total area of the disk and center it on a focal lattice cell. We then calculate the area of each cell that is covered by the disk and divide by the total area of the disk. These proportions become discrete probabilities of dispersal to a given cell relative to the original central cell. Once the probabilities are determined we used the alias method to build look-up tables from which the relative dispersal positions can be sampled in $O(1)$ time (Vose, 1991).

## 2.3 Analysis

Simulations were run for each of the four dispersal distributions under 9 levels of dispersal ($\sigma$ = 0.1, 0.25, 0.5, 1, 2, 4, 6, 8, and 10). These simulations were repeated for mutation rate $\mu = 10^{-4}$ and $\mu = 10^{-5}$ on two different landscapes: a 100 × 100 square landscape and a 500 × 20 nearly linear landscape. The simulations were run for 2 million generations after a burn-in period of 10,000 generations to allow the populations to reach an equilibrium. After the burn-in, data was collected every 1,000 generations to decrease correlations between samples. The populations were sampled by recording a transect of individuals in a single straight row across the landscape.

For each sampled transect, all possible pairs of individuals were placed into distance classes based on the minimum distance between them. The number of identical pairs was determined and recorded as a proportion of the total number of pairs in the distance class. The probabilities for each distance class were then averaged over all sampled generations.

Under isolation-by-distance, individuals geographically near one another will tend to be genetically similar, and this similarity will decrease as the distance between pairs of individuals increases. Therefore, isolation-by-distance is measured by constructing correlograms of genetic-similarity between individuals versus the distance between them. Genetic similarity can be measured using identity-by-descent, identity-by-state, relatedness, conditional kinship, or F-coefficients and can be based on coalescent times, an ancestral population, or the current population (Hardy and Vekemans, 1999; Hardy, 2003; Malécot, 1969; Rousset, 1997, 2002; Wang, 2014). For two-dimensional populations, genetic similarity is often plotted against the log-distance separating pairs because theory predicts that this relationship is linear (Barton *et al*, 2013; Hardy and Vekemans, 1999; Rousset, 2000).

Because the amount of identity-by-descent in each distance class depends on the mutation rate,

we calculated the kinship coefficient for each class, which is nearly independent of mutation rate:

$$F_{ij} = \frac{p_{ij} - \bar{p}}{1 - \bar{p}} \approx \frac{E[T] - E_{ij}[T]}{E[T]} \quad (1)$$

where $p_{ij}$ is the probability of identity-by-descent between haploid individuals $i$ and $j$ and $\bar{p}$ is the probability of identity-by-descent between random haploid individuals in the current sample (Hardy and Vekemans, 1999). The kinship coefficient is related to differences in the expected coalescent times between a specific pair of individuals and a random pair in the population (Barton *et al*, 2013). Kinship coefficients were calculated for each transect and then averaged across transects for each distance class.

For each simulation, we calculated the average number of unique alleles in a 100-individual transect ($\bar{k}$) and the average squared distance between parents and children ($2s^2$). Using $\bar{k}$, we estimated the population-level diversity, $\hat{\theta}_k$ (Ewens, 2004, eq. 9.32) and estimated effective haploid population size as $\hat{N}_e = \hat{\theta}_k / 2\mu$ and effective density as $\hat{D}_e = \hat{N}_e / A$, where $A = 10,000$. Finally, neighborhood size was estimated as $N_b = 4\pi s^2 \hat{D}_e$. While it is possible to estimate neighborhood size through regression (Barton *et al*, 2013; Hardy and Vekemans, 1999; Rousset, 2000), we found that it was more accurate to keep track of dispersal and calculate $N_b$ directly.

## 3 Results

Three simulations produced pathological output due to the inability of gametes to disperse away from their parents: Rayleigh and triangular with $\sigma = 0.1$ and triangular when $\sigma = 0.25$. For simplicity, we will not include these simulations in some of the analyses below. Furthermore, results for $\sigma > 4$ were often not qualitatively different than $\sigma = 4$, and we will elide them as well. Unless otherwise noted, all results were for mutation rate of $\mu = 10^{-4}$ and lattice size of $100 \times 100$. Additional results can be found in Supplementary Material.

### 3.1 Spatial Distribution of Haplotypes

Isolation-by-distance is expected to result in concentrated patches of identical alleles (Epperson, 1995). The genetic structure produced in our simulations was qualitatively consistent across distri-butions, and structure decreased as dispersal increased (Figure 1).

## 3.2 Allelic Diversity and Effective Population Sizes

The distribution of unique alleles is similar for each dispersal kernel. Differences in effective population size between simulations can be measured by comparing the number of unique alleles observed in the transects (Figure 2). Different dispersal kernels produce similar levels of diversity, while the effective population size appears to be noticeably bigger than the census size if $\sigma < 1$ (Table 2), in agreement with Maruyama (1972).

## 3.3 Spatial Autocorrelation and Isolation-by-Distance

To quantify the patterns of isolation-by-distance, the pairwise kinship coefficient (Eq. 1) was measured for each sampled population as a function of distance . All four dispersal kernels produced nearly identical patterns of isolation-by-distance when $\sigma \geq 0.5$ (Figure 3).

As predicted, neighborhood sizes were also similar among the different dispersal kernels (Table 2). The mean-squared parent-offspring dispersal distances were very close to the expectation. Most of the differences in neighborhood size is introduced through the estimation of the effective density.

## 3.4 Relative Execution Time of Dispersal Functions

To compare the run time for the different dispersal functions we ran simulations for each distribution for 1,000 generations on a 100 × 100 landscape. The CPU time was averaged over 5 different runs (Table 1e). The half-normal simulations had the longest process time, followed closely by the exponential simulations which ran in 99.6% the time of the half-normal. The Rayleigh distribution was modeled by drawing two axial distances from a normal distribution and adding each to the coordinates of the current position. This process avoided the expensive conversion from polar to Cartesian coordinates and ran in 34.2% the time of the half-normal. The triangular distribution function drew positions from a discrete distribution in constant time and also avoided conversion between coordinate systems. The process time for the discrete triangular distribution simulations was greatly reduced, running in 21.7% of the time of the half-normal distribution and 63.4% the time of the Rayleigh distribution.

## 4 Discussion

As predicted, the different dispersal kernels show similar patterns of isolation-by-distance and spatial structure. The similarity between the kernels is consistent across different landscapes, boundary conditions, and mutation rates confirming that the genetic consequences of dispersal are highly dependent on $\sigma^2$.

When $\sigma < 0.5$, the different dispersal distributions behave differently. These differences may be attributed to the discrete nature of the simulation landscape. Approximating continuous dispersal on a discrete lattice will introduce obvious biases when the dispersal distance is small compared to the scale of the lattice nodes (Chipperfield *et al*, 2011). Here the distance between nodes is one lattice unit so dispersal has to exceed at least a distance of 0.5 lattice units to leave the original cell. At the lowest dispersal level, $\sigma = 0.1$, the theoretical probabilities of dispersal exceeding 0.5 are 0.0067, 0.0048, $3.7 \times 10^{-6}$ and 0 for the exponential, half-normal, Rayleigh, and triangular kernels respectively. The simulation results seem to follow this pattern with the exponential simulations showing the most dispersal followed by the half-normal and the Rayleigh. When $\sigma$ is increased to 0.25 the first three distributions begin to show similar patterns of isolation-by-distance.

We constructed efficient sampling algorithms for our dispersal kernels, allowing us to make objective comparisons about their computational utility. The Rayleigh, exponential, and half-normal all used a ziggurat algorithm (Marsaglia and Tsang, 2000) which is the most efficient technique currently available for them. For the Rayleigh kernel, we drew the dispersal location based on two independent normal distributions, while the exponential and half-normal samples required conversion from polar to Cartesian coordinates. This difference in complexity is reflected in the relative execution times (Table 1c). We were able to simulate from the triangular distribution the fastest, roughly 50% faster than the Rayleigh distribution, which is itself efficient. This was achieved by discretizing dispersal by calculating the proportion of a cell that was contained in a circle or radius $2\sigma$. This discrete distribution was then simulated from in $O(1)$ time using the alias method (Vose, 1991). As a result, a triangular pattern of dispersal can be generated from a single uniform pseudo-random number, using only integer arithmetic.

The triangular distribution would seem to be an odd choice for modeling dispersal, especially because no dispersal can happen if $\sigma \leq 0.25$. However, we have demonstrated that as long as the proper scale is used, this distribution produces the same patterns of isolation-by-distance as distributions that are typically used. Furthermore, we have also demonstrated that simulating dispersal simulating dispersal from a discretized triangular distribution is much faster than the other dispersal distributions. Because speed is an important factor in deploying isolation-by-distance simulations in analytical contexts, e.g. approximate Bayesian computation, we recommend the triangular distribution be used. See the appendix for descriptions of how to use the triangular distribution in simulations.

In addition to the obvious computational advantage, the triangular distribution is interesting from a theoretical standpoint. In two-dimensions, it produces a uniform distribution over a circle with area $4\pi\sigma^2$, which perfectly defines the area of Wright's neighborhood, $N_b = 4\pi\sigma^2 D$. This suggests that the triangular distribution perfectly defines a local panmictic unit where parents of central individuals may be treated as if drawn at random. In that sense, the triangular distribution is an obvious candidate for a null dispersal model in isolation-by-distance studies.

## 5 Data Accessibility

Source code for simulations can be found at https://github.com/tfursten/IBD. Simulation parameters and results can be accessed on DRYAD doi:10.5061/dryad.jg7vn or at http://ra.cartwrig.ht/data/furstenau2015a-ibd.tar.bz2.

## 6 Acknowledgments

The authors would like to thank R. Schwartz and D. Winter for helpful comments, and K. Dai for programming tips.

# A  Xorshift Random Number Generator

Xorshift is a type of pseudo-random number generator that relies on exclusive-or and bitshift operators (Marsaglia, 2003). Xorshift is one of the most efficient, high-quality random-number generators known. Our implementation is a 64-bit xorshift with shift parameters (5, 15, 27) added to a Weyl series to decrease bit correlations (Brent, 2007). It passes the BigCrush tests in the TestU01 suite (L'Ecuyer and Simard, 2007).

# B  Generating from a Triangular Distribution

Inverse sampling can be used to generate values from a triangular distribution. — Note that we are only working with monotonically increasing triangular distributions and not more general formulations. — If $u$ is uniformly distributed in $(0, 1)$, the value $d = 2s\sqrt{u}$ has a triangular distribution with parameter $s$. However, a modified rejection sampling algorithm is faster. If $u_1$ and $u_2$ are independent and uniformly distributed in $(0, 1)$, then $d = 2s \max(u_1, u_2)$ also has a triangular distribution. Because we can generate 32-bit values for both $u_1$ and $u_2$ from a single 64-bit random number, this second algorithm is more efficient than the first. While it is possible to construct a ziggurat algorithm (Marsaglia and Tsang, 2000) for a triangular distribution, our second algorithm is more efficient because it involves fewer steps and never rejects.

We compared the speed of these algorithms and a naive rejection sampler using the medium Crush tests (L'Ecuyer and Simard, 2007). This allowed us to compare the speeds of these algorithms in a data-intensive application as well as verify that the algorithms produced independent and identically distributed values from the correct distribution. The 'maximum' algorithm took 1656 seconds to complete, while the 'sqrt' took 1700s and the rejection sampler took 1911s. The maximum algorithm produced faster execution, but only sped up the tests by 3% over sqrt.

# C Generating Discrete Two-Dimensional Dispersal from a Triangular Distribution

We can use the maximum algorithm above to generate the values in polar coordinates and convert them to Cartesian coordinates; however, this requires calculating sine and cosine functions, which we would rather not do. When modeling dispersal on a lattice, the bounded nature of the triangular distribution allows dispersal to be modeled discretely. To discretize this distribution on a rectangular lattice we must determine the probabilities for each cell which are proportional to the area of the cell that is covered by a disk of radius $r = 2\sigma$ (centered on a focal cell). The algorithm described here produces probability tables by calculating the appropriate area for each cell and dividing by the total area. We assume that cells are squares with unit area.

Since the disk is symmetrical, this algorithm may be simplified by calculating areas for quadrant I of the disk and mirroring those values to the other quadrants. We further simplify by calculating approximately half of the areas for quadrant I and mirroring those as well. — Note that this results in cells along the x and y axes having an area of 1/2. — Starting at the center of the focal cell ($y_0 = 0$), we record the top/bottom boundary of each cell along the y-axis up to the radius: $y_1 = 0.5, y_2 = 1.5, \ldots, y_n = n - 0.5$ where $n = \sup_{n \in \mathbb{Z}} y_n \leq r$.

Next we calculate the area of the first column of cells which has a left boundary at $x_0 = 0$ and a right boundary at $x_1 = \min(0.5, r)$:

$$A = \int_{x_0}^{x_1} \sqrt{r^2 - x^x} \, dx$$

Starting with the bottom cell, we check if the area of a cell is less than the area of the column. If so, the cell is completely contained in the disk, and the cell is assigned a weight equal to its area. Its area is then subtracted from the area of the column. We continue this procedure until the the area of last cell is less than the remaining area of the column and assign the final cell a weight equal to the remaining area in the column.

We then move to the next column by setting $x_0 = 0.5$ and $x_1 = 1.5$. However, before we calculate the area, we must check if the edge of the disk passes through the bottom of the top cell. This occurs if $x_1^2 + y^2 > r^2$, where $y$ is the value of the bottom boundary of the cell. When this occurs, we split the column into two smaller columns and each column is processed just like before. We continue calculating the area of subsequent columns until we reach the column that contains the point $\{x, y\} = \{r/\sqrt{2}, r/\sqrt{2}\}$, which marks the point where the edge of the disk intersects the diagonal. After this column is processed, the weights for these cells can be copied symmetrically. The weight of each cell is divided by the total area of the disk and becomes a probability. These probabilities are then copied symmetrically to the other three quadrants. The completed table of probabilities can then be passed into the alias algorithm for discrete sampling (Vose, 1991).

Our implementation of a discretized triangular kernel can be found in src/disk.h and src/disk.cpp in the source code. Code for generating an alias table can be found in src/aliastable.h.

**Table 1:** ***Dispersal Kernels.*** *(a) The dispersal function and range. (b) Probability density function for $\sigma = 1$. (c) Relative execution time of simulations run for 1000 generations.*

| | Rayleigh | Exponential | Half-Normal | Triangular |
|---|---|---|---|---|
| **a** | $f(d, \alpha; \sigma) = \frac{1}{2\pi} \frac{d}{\sigma^2} e^{\frac{-d^2}{2\sigma^2}}$ $d \geq 0$ | $f(d, \alpha; \sigma) = \frac{1}{2\pi} \frac{1}{\sigma} e^{\frac{-d}{\sigma}}$ $d \geq 0$ | $f(d, \alpha; \sigma) = \frac{1}{2\pi} \frac{1}{\sigma\sqrt{\pi}} e^{\frac{-d^2}{4\sigma^2}}$ $d \geq 0$ | $f(d, \alpha; \sigma) = \frac{1}{2\pi} \frac{d}{2\sigma^2}$ $0 \leq d \leq 2\sigma$ |
| **b** | 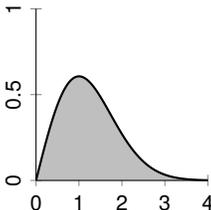 | 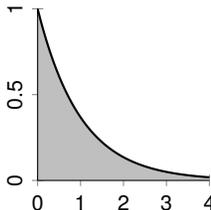 | 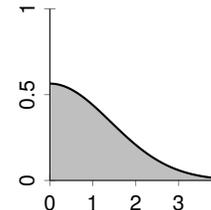 | 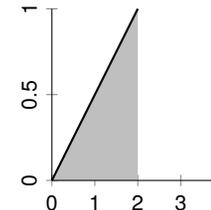 |
| **c** | 0.342 | 0.996 | 1 | 0.217 |

Table 2: *Estimated neighborhood sizes are similar across all dispersal distributions.* Estimates of the average number of alleles, $\bar{k}$, allele diversity, $\hat{\theta}_k$, effective population size, $\hat{N}_e$, effective population density, $\hat{D}_e$, dispersal, $s^2$, and neighborhood size, $\hat{N}_b$.

| | $\sigma$ | | | | | | | | | | | | | | | |
|---|---|---|---|---|---|---|---|---|---|---|---|---|---|---|---|---|
| | 0.25 | | | | 0.5 | | | | 1 | | | | 2 | | | |
| | Ray | Exp | Nor | Tri | Ray | Exp | Nor | Tri | Ray | Exp | Nor | Tri | Ray | Exp | Nor | Tri |
| $\bar{k}$ | 17.4 | 16.3 | 15.8 | 100.0 | 9.6 | 10.9 | 10.2 | 9.0 | 8.2 | 9.0 | 8.5 | 8.1 | 8.3 | 8.5 | 8.3 | 8.2 |
| $\hat{\theta}_k$ | 5.8 | 5.3 | 5.0 | - | 2.4 | 2.9 | 2.6 | 2.2 | 1.9 | 2.2 | 2.0 | 1.9 | 2.0 | 2.0 | 2.0 | 1.9 |
| $\hat{N}_e$ | 29216 | 26416 | 25143 | - | 12101 | 14620 | 13213 | 10964 | 9669 | 11103 | 10136 | 9450 | 9757 | 10212 | 9827 | 9595 |
| $\hat{D}_e$ | 2.9 | 2.6 | 2.5 | - | 1.2 | 1.5 | 1.3 | 1.1 | 1.0 | 1.1 | 1.0 | 0.9 | 1.0 | 1.0 | 1.0 | 1.0 |
| $s^2$ | 0.043 | 0.061 | 0.059 | - | 0.32 | 0.27 | 0.29 | 0.39 | 1.08 | 1.04 | 1.05 | 1.12 | 4.10 | 4.08 | 4.06 | 4.11 |
| $\hat{N}_b$ | 1.6 | 2.0 | 1.9 | - | 4.9 | 5.0 | 4.8 | 5.4 | 13.1 | 14.5 | 13.4 | 13.3 | 50.3 | 52.4 | 50.1 | 49.6 |
| | 4 | | | | 6 | | | | 8 | | | | 10 | | | |
| | Ray | Exp | Nor | Tri | Ray | Exp | Nor | Tri | Ray | Exp | Nor | Tri | Ray | Exp | Nor | Tri |
| $\bar{k}$ | 8.6 | 8.8 | 8.5 | 8.6 | 8.6 | 8.7 | 8.7 | 8.8 | 8.7 | 8.6 | 8.6 | 8.9 | 8.8 | 8.7 | 8.7 | 8.7 |
| $\hat{\theta}_k$ | 2.1 | 2.1 | 2.1 | 2.1 | 2.1 | 2.1 | 2.1 | 2.1 | 2.1 | 2.1 | 2.1 | 2.2 | 2.1 | 2.1 | 2.1 | 2.1 |
| $\hat{N}_e$ | 10428 | 10670 | 10254 | 10427 | 10400 | 10603 | 10597 | 10639 | 10530 | 10427 | 10340 | 10788 | 10621 | 10576 | 10536 | 10528 |
| $\hat{D}_e$ | 1.0 | 1.1 | 1.0 | 1.0 | 1.0 | 1.1 | 1.1 | 1.1 | 1.1 | 1.0 | 1.0 | 1.1 | 1.1 | 1.1 | 1.1 | 1.1 |
| $s^2$ | 16.06 | 15.95 | 16.11 | 16.12 | 36.19 | 35.99 | 36.10 | 36.14 | 64.05 | 63.78 | 64.11 | 64.04 | 99.95 | 96.19 | 100.45 | 100.30 |
| $\hat{N}_b$ | 210.5 | 213.9 | 207.6 | 211.2 | 473.0 | 479.5 | 480.7 | 483.1 | 847.5 | 835.7 | 833.0 | 868.1 | 1334.0 | 1278.3 | 1329.9 | 1326.9 |

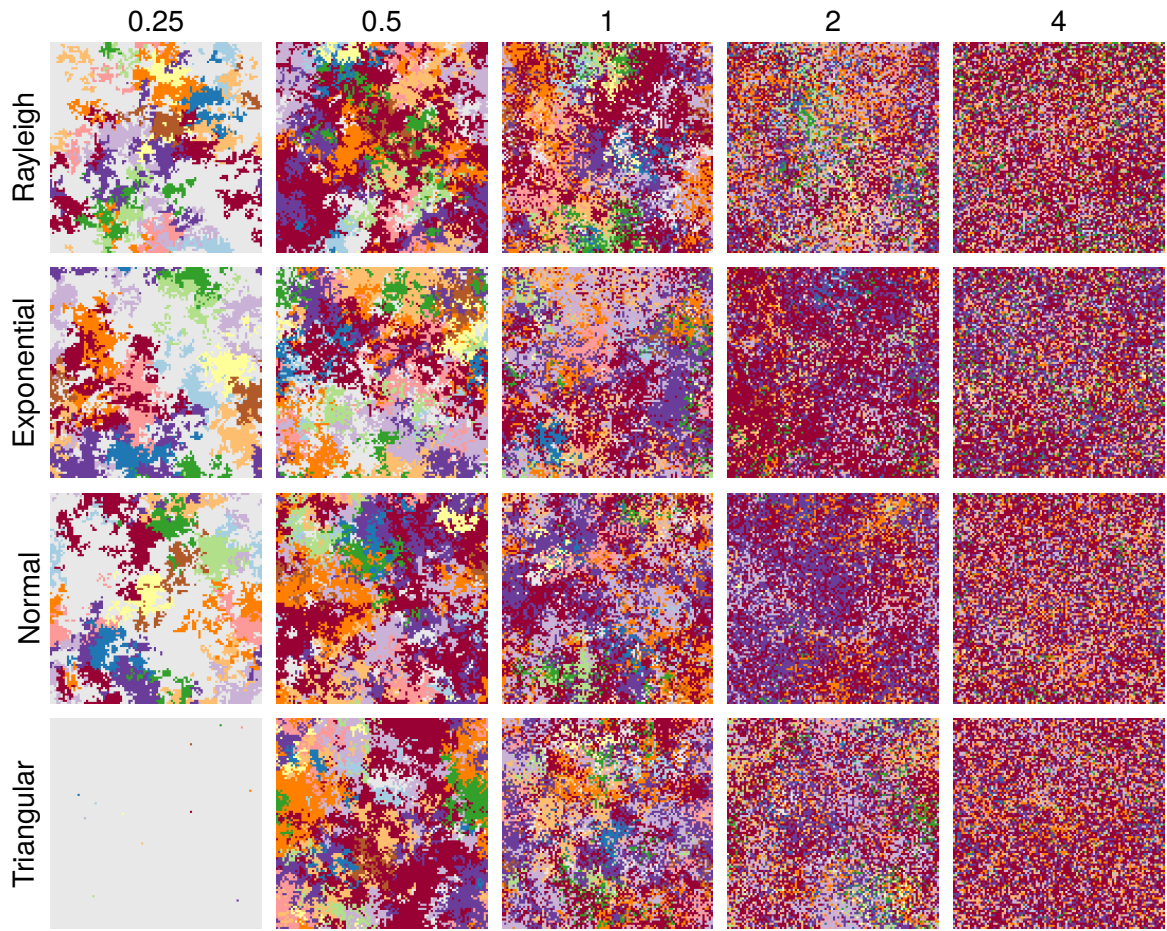

**Figure 1:** ***The spatial distribution of haplotypes is qualitatively similar for each dispersal kernel.*** *Each box depicts a different simulation at 50,000 generations after a 10,000 generation burn-in period using the dispersal distribution and $\sigma$ parameter specified. The darker shaded pixels represent the top 12 high frequency alleles and the light-gray pixels represent the rest of the alleles in the population.*

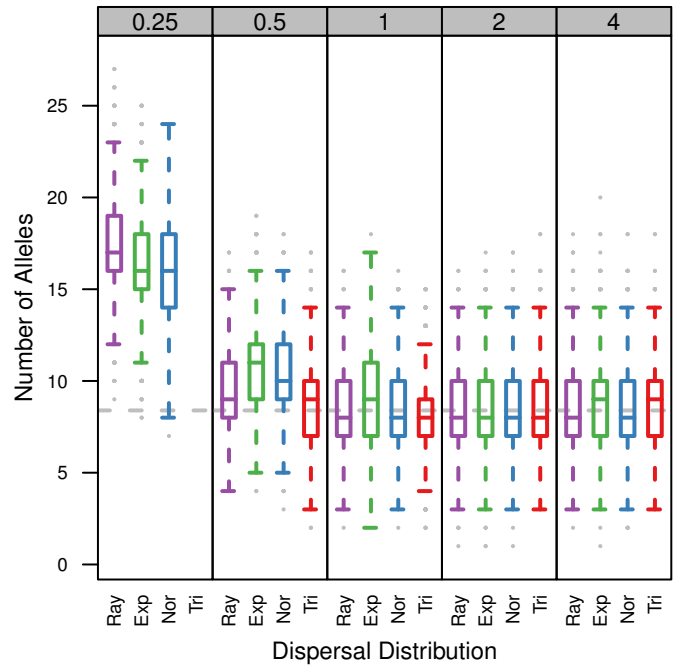

**Figure 2:** *The distribution of unique alleles is similar for each dispersal kernel.* Each panel represents a $\sigma$ and contains box-whisker plots summarizing the number of unique alleles (k) found in 2000 sampled transects. The top and bottom of the boxes represent the 75% and 25% quartiles, while the central bar represents the median. The gray dots outside the whiskers represent outliers. The dashed horizontal line represents the expectation under the infinite alleles model.

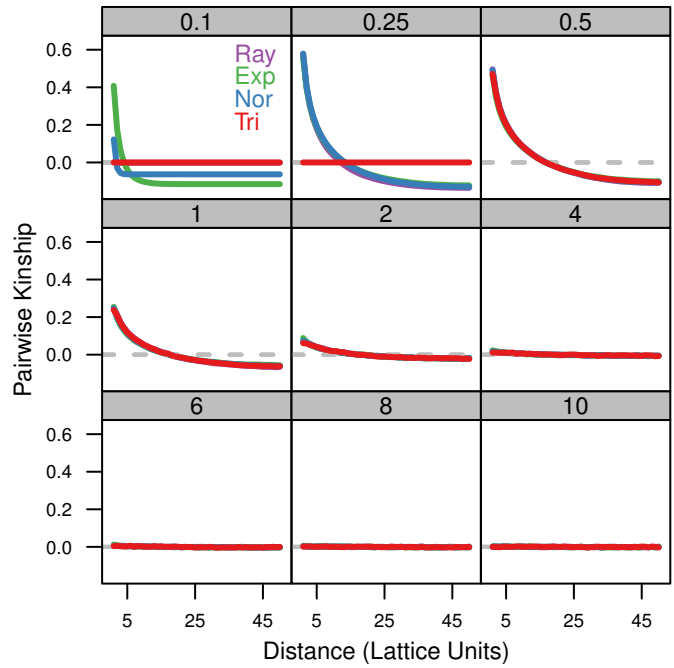

**Figure 3:** *Isolation-by-distance is consistent across dispersal kernels.* *Average pairwise kinship coefficients ($F_{ij}$) are shown for distance classes 1-50 and for each distribution and $\sigma$. Positive values indicate that individuals in a distance class have an excess of identical genes while negative values indicate a deficit compared to the rest of the sample. While there are four lines plotted for each panel, they usually cannot be distinguished due to overlap.*

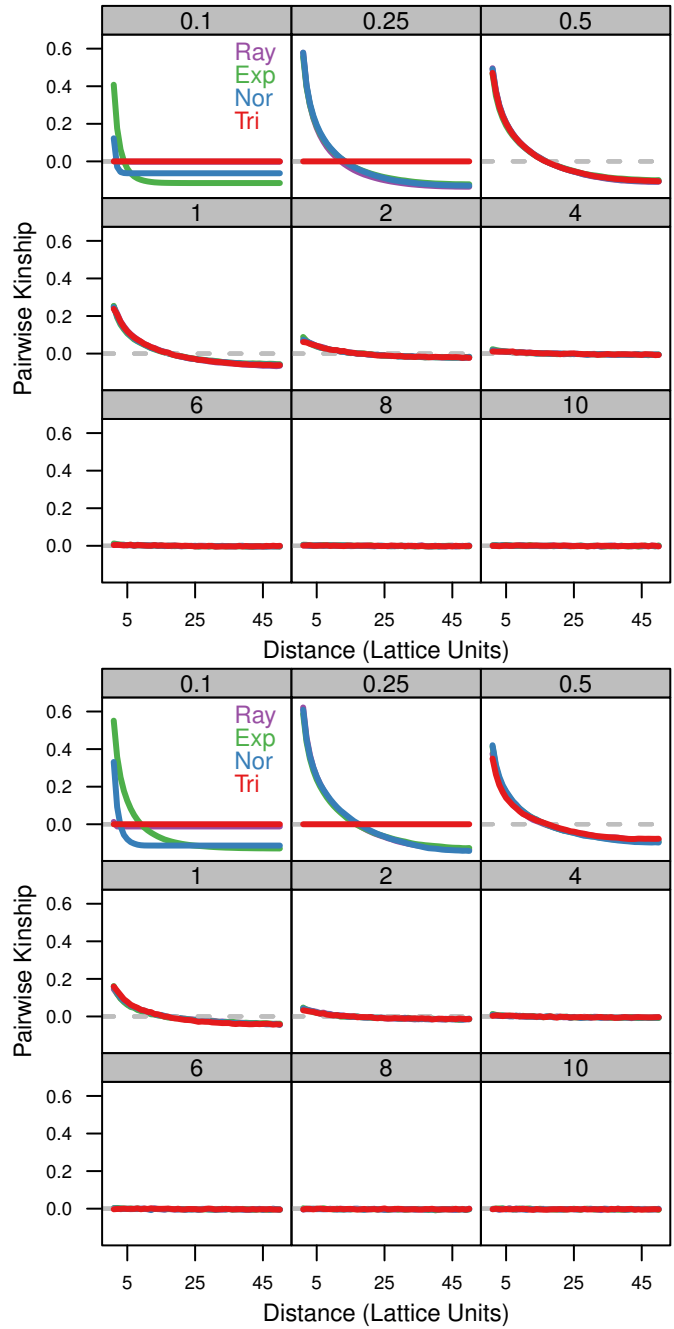

**Figure 1:** *Isolation-by-distance is consistent across dispersal kernels for different mutation rates.* Simulations were run on a $100 \times 100$ torus with $\mu = 10^{-4}$ (top: same as Figure 3 in main) and $10^{-5}$ (bottom). Average pairwise kinship coefficients ($F_{ij}$) are shown for distance classes 1-50 for each distribution and $\sigma$.

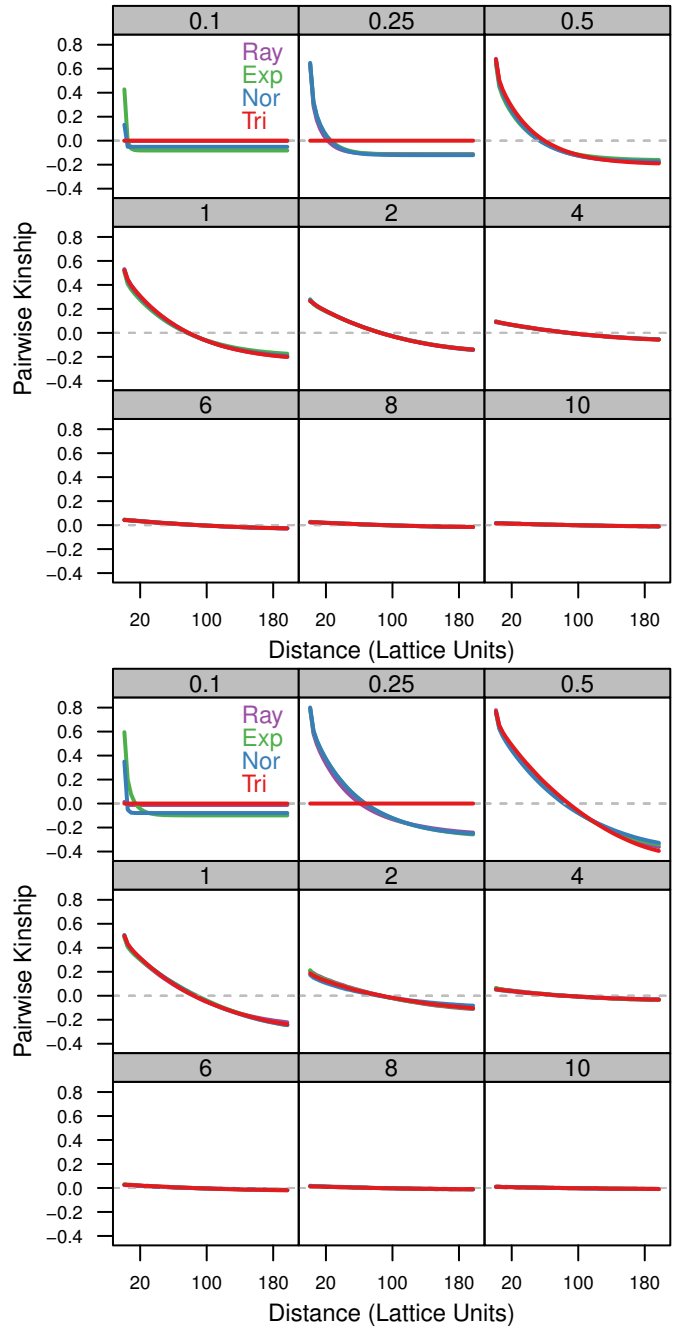

**Figure 2:** *Isolation-by-distance is consistent across dispersal kernels for different landscapes.* Simulations were run on a $20 \times 500$ torus with $\mu = 10^{-4}$ (top) and $10^{-5}$ (bottom). Average pairwise kinship coefficients ($F_{ij}$) are shown for distance classes 1-200 for each distribution and $\sigma$.

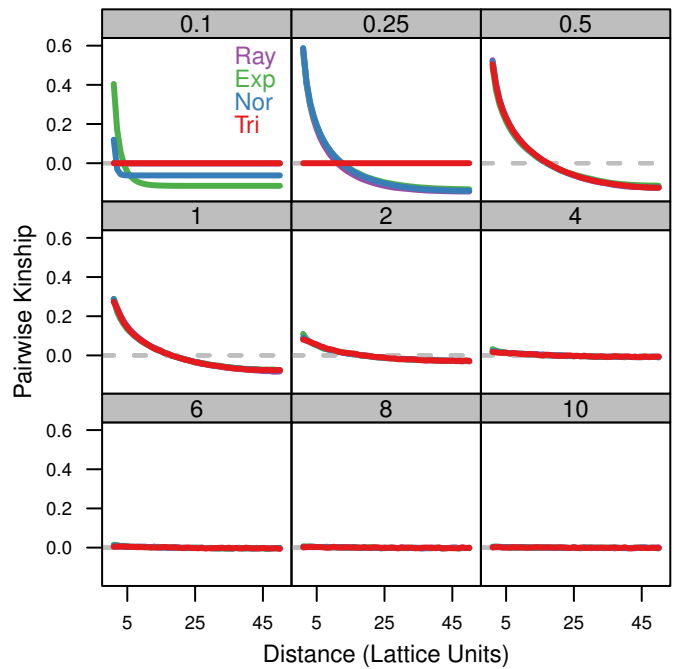

**Figure 3:** *Isolation-by-distance is consistent across dispersal kernels for different boundary conditions.* Simulations were run on a $100 \times 100$ landscape with absorbing boundaries and $\mu = 10^{-4}$. Average pairwise kinship coefficients ($F_{ij}$) are shown for distance classes 1-50 for each distribution and $\sigma$.

**Table 1:** *Observed dispersal is similar to expected values across all dispersal distributions when $\sigma \geq 0.5$.* Estimates of the average number of alleles, $\bar{k}$, allele diversity, $\hat{\theta}_k$, effective population size, $\hat{N}_e$, effective population density, $\hat{D}_e$, dispersal, $s^2$, and neighborhood size, $\hat{N}_b$. The statistics were calculated as described in main text. For $\sigma = 0.1$, the observed $s^2$ for each distribution is much different than the expected value of 0.01. This is due to the descritization of continuous distributions on a discrete lattice (see main text for more discussion). At $\sigma = 0.1$, the theoretical probabilities that dispersal will exceed a distance of 0.5 (minimum distance required to disperse out of original cell) are 0.0067, 0.0048, $3.7 \times 10^{-6}$ and 0 for the exponential, half-normal, Rayleigh and triangular distributions, respectively. At $\sigma = 0.25$ the probability of dispersal for the triangular distribution is still zero but the other distributions show dispersal near the expected $\sigma^2 = 0.06$.

|  | $\sigma$ | | | | | | | | | | | |
|---|---|---|---|---|---|---|---|---|---|---|---|---|
|  | 0.1 | | | | 0.25 | | | | .5 | | | |
|  | Ray | Exp | Nor | Tri | Ray | Exp | Nor | Tri | Ray | Exp | Nor | Tri |
| $\bar{k}$ | 99.77 | 43.1 | 81.6 | 100.0 | 17.4 | 16.3 | 15.8 | 100.0 | 9.6 | 10.9 | 10.2 | 9.0 |
| $\hat{\theta}_k$ | - | 28.23 | 205.37 | - | 5.8 | 5.3 | 5.0 | - | 2.4 | 2.9 | 2.6 | 2.2 |
| $\hat{N}_e$ | - | 141152 | 1026851 | - | 29216 | 26416 | 25143 | - | 12101 | 14620 | 13213 | 10964 |
| $\hat{D}_e$ | - | 14.12 | 102.69 | - | 2.9 | 2.6 | 2.5 | - | 1.2 | 1.5 | 1.3 | 1.1 |
| $s^2$ | - | 0.00194 | 0.00008 | - | 0.043 | 0.061 | 0.059 | - | 0.32 | 0.27 | 0.29 | 0.39 |
| $\hat{N}_b$ | - | 0.3 | 0.1 | - | 1.6 | 2.0 | 1.9 | - | 4.9 | 5.0 | 4.8 | 5.4 |

|  | 1 | | | | 2 | | | | 4 | | | |
|---|---|---|---|---|---|---|---|---|---|---|---|---|
|  | Ray | Exp | Nor | Tri | Ray | Exp | Nor | Tri | Ray | Exp | Nor | Tri |
| $\bar{k}$ | 8.2 | 9.0 | 8.5 | 8.1 | 8.3 | 8.5 | 8.3 | 8.2 | 8.6 | 8.8 | 8.5 | 8.6 |
| $\hat{\theta}_k$ | 1.9 | 2.2 | 2.0 | 1.9 | 2.0 | 2.0 | 2.0 | 1.9 | 2.1 | 2.1 | 2.1 | 2.1 |
| $\hat{N}_e$ | 9669 | 11103 | 10136 | 9450 | 9757 | 10212 | 9827 | 9595 | 10428 | 10670 | 10254 | 10427 |
| $\hat{D}_e$ | 1.0 | 1.1 | 1.0 | 0.9 | 1.0 | 1.0 | 1.0 | 1.0 | 1.0 | 1.1 | 1.0 | 1.0 |
| $s^2$ | 1.08 | 1.04 | 1.05 | 1.12 | 4.10 | 4.08 | 4.06 | 4.11 | 16.06 | 15.95 | 16.11 | 16.12 |
| $\hat{N}_b$ | 13.1 | 14.5 | 13.4 | 13.3 | 50.3 | 52.4 | 50.1 | 49.6 | 210.5 | 213.9 | 207.6 | 211.2 |

|  | 6 | | | | 8 | | | | 10 | | | |
|---|---|---|---|---|---|---|---|---|---|---|---|---|
|  | Ray | Exp | Nor | Tri | Ray | Exp | Nor | Tri | Ray | Exp | Nor | Tri |
| $\bar{k}$ | 8.6 | 8.7 | 8.7 | 8.8 | 8.7 | 8.6 | 8.6 | 8.9 | 8.8 | 8.7 | 8.7 | 8.7 |
| $\hat{\theta}_k$ | 2.1 | 2.1 | 2.1 | 2.1 | 2.1 | 2.1 | 2.1 | 2.2 | 2.1 | 2.1 | 2.1 | 2.1 |
| $\hat{N}_e$ | 10400 | 10603 | 10597 | 10639 | 10530 | 10427 | 10340 | 10788 | 10621 | 10576 | 10536 | 10528 |
| $\hat{D}_e$ | 1.0 | 1.1 | 1.1 | 1.1 | 1.1 | 1.0 | 1.0 | 1.1 | 1.1 | 1.1 | 1.1 | 1.1 |
| $s^2$ | 36.19 | 35.99 | 36.10 | 36.14 | 64.05 | 63.78 | 64.11 | 64.04 | 99.95 | 96.19 | 100.45 | 100.30 |
| $\hat{N}_b$ | 473.0 | 479.5 | 480.7 | 483.1 | 847.5 | 835.7 | 833.0 | 868.1 | 1334.0 | 1278.3 | 1329.9 | 1326.9 |